\documentstyle[twoside,fleqn,espcrc2,epsfig]{article}
\newcommand{\qsq    }{\mbox{$Q^2$}}
\newcommand{\psq    }{\mbox{$P^2$}}
\newcommand{\qzm    }{\mbox{$\langle \qsq \rangle$}}
\newcommand{\pzm    }{\mbox{$\langle \psq \rangle$}}
\newcommand{\gev    }{\mbox{$\rm GeV$}}
\newcommand{\gevsq  }{\mbox{$\rm GeV^2$}}
\newcommand{\der    }{\mbox{${\mathrm d}$}}
\newcommand{\dsigdx }{\mbox{$\der\sigma/\der x$}}
\newcommand{\ftqed  }{\mbox{$F_{\mathrm{2,QED}}^{\gamma}$}}
\newcommand{\faqed  }{\mbox{$F_{\mathrm{A,QED}}^{\gamma}$}}
\newcommand{\fbqed  }{\mbox{$F_{\mathrm{B,QED}}^{\gamma}$}}
\newcommand{\ft     }{\mbox{$F_{2}^{\gamma}$}}
\newcommand{\stt    }{\mbox{$\sigma_{\mathrm{TT}}$}}
\newcommand{\slt    }{\mbox{$\sigma_{\mathrm{LT}}$}}
\newcommand{\stl    }{\mbox{$\sigma_{\mathrm{TL}}$}}
\newcommand{\sll    }{\mbox{$\sigma_{\mathrm{LL}}$}}
\newcommand{\ttt    }{\mbox{$\tau_{\mathrm TT}$}}
\newcommand{\ttl    }{\mbox{$\tau_{\mathrm TL}$}}
\newcommand{\barph  }{\mbox{$\bar{\phi}$}}
\newcommand{\cosph  }{\mbox{$\cos\barph$}}
\newcommand{\costph }{\mbox{$\cos 2\barph$}}
\newcommand{\mumu   }{\mbox{$\mu\mu$}}
\newcommand{\az     }{\mbox{$\chi$}}
\newcommand{\aem    }{\mbox{$\alpha$}}
\newcommand{\aemsq  }{\mbox{$\aem^2$}}
\newcommand{\faoft  }{\mbox{$\faqed/\ftqed$}}
\newcommand{\fboft  }{\mbox{$\fbqed/\ftqed$}}
\newcommand{\chidof }{\mbox{$\chi^2/\mathrm{dof}$}}

\newcommand{\Y      }[3]{\mbox{$#1\,^{+\,#2}_{-\,#3}$}}
\newcommand{\chiq   }{\mbox{$\chi^2$}}
\newcommand{\omfwq  }{\mbox{${\mathcal{O}}({m_\mu^2}/{W^2})$}}
\newcommand{\ee     }{\mbox{${\mathrm e}{\mathrm e}$}}

\newcommand{\ftxq   }{\mbox{$\ft(x,\qsq)$}}
\newcommand{\flxq   }{\mbox{$\fl(x,\qsq)$}}
\newcommand{\feff   }{\mbox{$F_{\mathrm{eff}}^{\gamma}$}}

\newcommand{\flow   }{\mbox{$1/N~{\rm d}E/{\rm d}\eta$}}

\newcommand{\Wvis   }{\mbox{$W_{\mathrm{vis}}$}}

\newcommand{\kt     }{\mbox{$k_{\mathrm t}$}}
\newcommand{\ktsq   }{\mbox{$k_{\mathrm t}^2$}}

\newcommand{\kn     }{\mbox{$k_0$}}

\newcommand{\epem   }{\mbox{$\mathrm{e}^+\mathrm{e}^-$}}
\newcommand{\ftp    }{\mbox{$F_{\mathrm{2}}^{\mathrm{p}}$}}
\newcommand{\lsim   }{\mbox{$\raisebox{-0.5mm}{$\stackrel{<}{\scriptstyle{\sim}}$}$}}

\newcommand{\fti    }{\mbox{$F_{2,i}^{\gamma}$}}

\newcommand{\ftc    }{\mbox{$F_{2,\mathrm{c}}^{\gamma}$}}
\newcommand{\almz   }{\mbox{$\alpha_{s}(M^{2}_z)$}}
\newcommand{\qnsq   }{\mbox{$Q_0^{2}$}}
\newcommand{\wsq    }{\mbox{$W^{2}$}}
\newcommand{\ssee   }{\mbox{$\sqrt{s_{\rm e e}}$}}

\newcommand{\fl     }{\mbox{$F_{\mathrm{L}}^{\gamma}$}}

\title{The Structure of Quasi-Real and Virtual Photons
\thanks{Invited talk given at the PHOTON 99 Conference, Freiburg, 
        Germany, May 1999, to appear in the Proceedings.}
}
\author{Richard Nisius, \it{CERN, CH-1211 Gen\`eve 23, Switzerland,
        e-mail: Richard.Nisius@cern.ch}
       }
\begin{document}
%
%
\begin{abstract}
 This review covers the measurements of the QED and QCD structure 
 of quasi-real and virtual photons from the reaction
 $\ee \rightarrow \ee \gamma^\star\gamma^{(\star)}\rightarrow \ee X$, 
 and is an update of the discussion presented in~\cite{NISIUS}.
\end{abstract}
%
%
\maketitle
%
%
\section{Introduction}
\label{sec:intro}
\begin{figure}[htb]
\begin{center}
\mbox{}\vspace{-1.3cm}
{\includegraphics[width=1.0\linewidth]{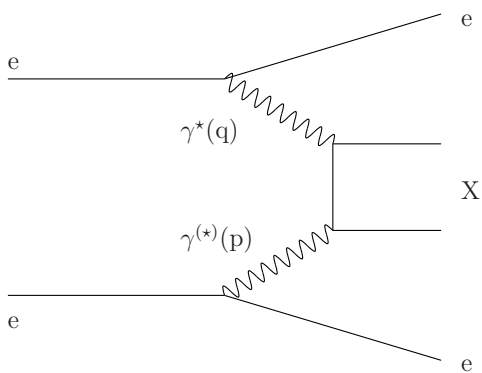}}
\mbox{}\vspace{-1.3cm}
\caption{
         A diagram of the reaction $\ee \rightarrow \ee X$,
         proceeding via the exchange of two photons.
        }\label{fig01}
\mbox{}\vspace{-1.3cm}
\end{center}
\end{figure}

 One of the most powerful tools to investigate the structure of quasi-real
 photons, $\gamma$, is the measurement of photon structure functions
 in deep inelastic electron-photon scattering at electron-positron
 colliders, shown in Figure~\ref{fig01}.
 \par
%
%
 These measurements have by now a tradition of eighteen years
 since the first \ft\ was obtained by PLUTO~\cite{PLU-8102}. 
 The LEP accelerator is a unique place for the measurements of 
 photon structure functions until a high energy linear collider is realised.
 It is unique because of the large coverage in \qsq\ owing to the various
 beam energies covered within the LEP programme, and 
 due to the high luminosities delivered to the experiments.
 The main idea is that by measuring the differential cross-section
 \newline
%
 \begin{eqnarray}
     \frac{d^2\sigma_{{\rm e}\gamma\rightarrow{\rm e}{\rm X}}}{dxdQ^2}
 &=& \frac{2\pi\aemsq}{x\,Q^{4}}
     \left[\left( 1+(1-y)^2\right) \ftxq \right.\nonumber\\
 & & \left. \quad\quad\quad - y^{2} \flxq\right]\, ,
 \end{eqnarray}
%
 one obtains the photon structure function \ft.
 Here $\qsq=-q^2$ is the absolute value of the four momentum squared 
 of the virtual photon, $\gamma^{\star}$, $x$ and $y$ are the usual 
 dimensionless variables of 
 deep inelastic scattering and \aem\ is the fine structure constant.
 In the region of small $y$ studied ($y\ll 1$) the contribution 
 of the term proportional to \fl\ is small and it is usually neglected.
 In leading order \ft\ is proportional to the parton content,
 $f_{i,\gamma}$, of the photon,
 $\ft=x\sum_{c,f}\,e_q^2\,(f_{\rm q,\gamma}+f_{\bar{\rm q},\gamma})$ 
 for quarks, and therefore reveals the internal structure of the photon.
 \par
 Because the energy of the quasi-real photon is not known, 
 $x$ has to be derived by measuring the invariant mass of the
 final state $X$, which consists of \mumu\ pairs
 for the investigation of the QED structure of the photon,
 or of hadrons created by a quark pair in studies of \ft.
 In the case of \mumu\ final states the invariant mass can be
 determined accurately, and measurements of the QED structure
 are generally statistically limited.
 For hadronic final states the measurement of $x$  is a source of
 significant uncertainties which makes measurements of \ft\ mainly 
 systematics limited.
 \par
 If both photons are virtual an effective structure function
 \feff\ of virtual photons can be determined in the region 
 $\qsq\gg\psq\gg\Lambda^2$ and,
 for $\qsq\approx\psq\gg\Lambda^2$ the differential cross-section
 for the exchange of two virtual photons is probed.
 Here $\Lambda$ is the QCD scale.
 The different measurements performed to investigate the QED and 
 QCD structure of the photon are discussed in the following.
%
%
\section{The QED structure of the photon}
\label{sec:qedres}
 Several measurements concerning the QED structure of the 
 photon have been performed by various experiments.
 Prior to LEP, mainly \ftqed\ of quasi-real photons was measured.
 The LEP experiments refined the analysis of the \mumu\ final state, 
 and derived more information on the QED structure of the photon.
 The interest in the investigation of the QED structure of the photon
 is twofold.
 Firstly the investigations serve as tests of QED to order
 ${\mathcal{O}}(\aem^4)$, but secondly, and also very important,
 the investigations are used to refine the experimentalists tools in a real
 but clean experimental situation to investigate the possibilities of 
 extracting similar information from the much more complex hadronic final
 state.
%
%
\subsection{The structure function \ftqed}
\label{sec:qedresf2}
\begin{figure}[htb]
\begin{center}
\mbox{}\vspace{-0.7cm}
{\includegraphics[width=1.0\linewidth]{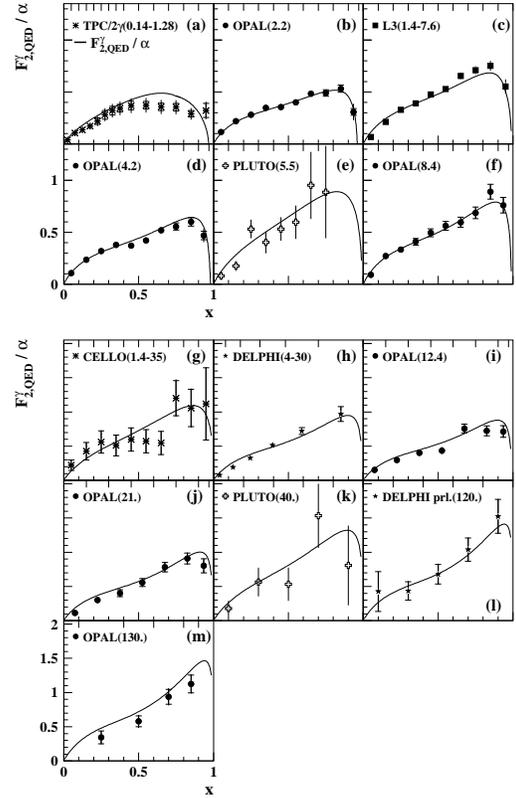}}
\mbox{}\vspace{-1.3cm}
\caption{
         A summary of \ftqed\ measurements.
         The quoted errors for h) are statistical only.
        }\label{fig02}
\mbox{}\vspace{-1.6cm}
\end{center}
\end{figure}

 The structure function \ftqed\ has been measured using data 
 in the approximate range of average virtualities \qzm\ of
 $0.45-130$~\gevsq.
 Results were published by
 CELLO~\cite{CEL-8301},
 DELPHI~\cite{DEL-9601},
 L3~\cite{L3C-9801},
 OPAL~\cite{OPALPR271},
 PLUTO~\cite{PLU-8501}
 and
 TPC/2$\gamma$~\cite{TPC-8401}.
 Additional preliminary results are available from
 ALEPH~\cite{BRE-9701} and DELPHI~\cite{ZIN-9901}.
 The ALEPH results are preliminary since two years and therefore 
 they are not considered here.
 The DELPHI result at $\qzm=12.5$~\gevsq\ is going to replace the published
 measurement, which will still be used here.
 Special care has to be taken when comparing the experimental results
 to the QED predictions, because slightly different quantities are
 derived by the experiments.
 Some of the experiments express their result
 as an average structure function, $\langle\ftqed(x,\qsq)\rangle$, measured
 within their experimental acceptance in \qsq, whereas the other experiments
 unfold their result as a structure function for an average
 \qsq\ value, $\ftqed(x,\qzm)$.
 Figure~\ref{fig02} shows a summary of the \ftqed\ measurements 
 compared either to $\langle\ftqed(x,\qsq)\rangle$, assuming a flat 
 acceptance in \qsq, or to $\ftqed(x,\qzm)$, while
 using the appropriate values for \qsq\ and \qzm\ given by the experiments.
 For the measurements which quote an average virtuality \pzm\ of 
 the quasi-real photon for their dataset,
 this value is chosen in the comparison, otherwise $\psq=0$ is used.
 \par
%
\begin{figure}[htb]\unitlength 1pt
\begin{center}
\mbox{}\vspace{-1.0cm}
{\includegraphics[width=1.0\linewidth]{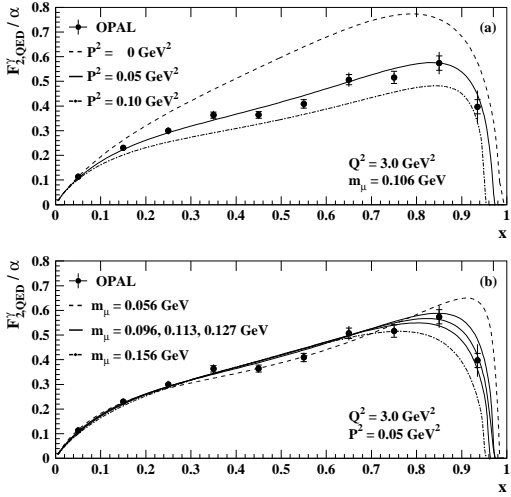}}
\mbox{}\vspace{-1.3cm}
\caption{
         The dependence of \ftqed\ on \psq\ and on the mass of the muon.
        }\label{fig03}
\mbox{}\vspace{-1.3cm}
\end{center}
\end{figure}

%
 There is agreement between the data and the QED expectations
 for about three orders of magnitude in \qsq.
 Some differences are seen for the  TPC/2$\gamma$ result, but at these
 low values of \qsq\ this could also be due to the simple averaging 
 procedure used for the theoretical prediction.
 The LEP data are so precise that the effect of the small virtuality
 of the quasi-real photon can clearly be established,
 as shown, for example, in Figure~\ref{fig03} for the most precise data
 from OPAL.
%
%
 The data are compared to the QED predictions of 
 \ftqed(x,\qzm,\pzm,$m_\mu$), where either \pzm\ or $m_\mu$ is varied.
 The mass of the muon is found to be $m_\mu=\Y{0.113}{0.014}{0.017}$~\gev,
 assuming the \pzm\ value predicted by QED.
 Although this is not a precise measurement of the mass of the muon it
 can serve as an indication on the achievable precision for the determination
 of $\Lambda$, if it only were for the pointlike contribution to \ft.
%
%
\subsection{Azimuthal correlations}
\label{sec:qedresaz}
\begin{figure}[htb]\unitlength 1pt
\begin{center}
\mbox{}\vspace{-1.2cm}
{\includegraphics[width=1.0\linewidth]{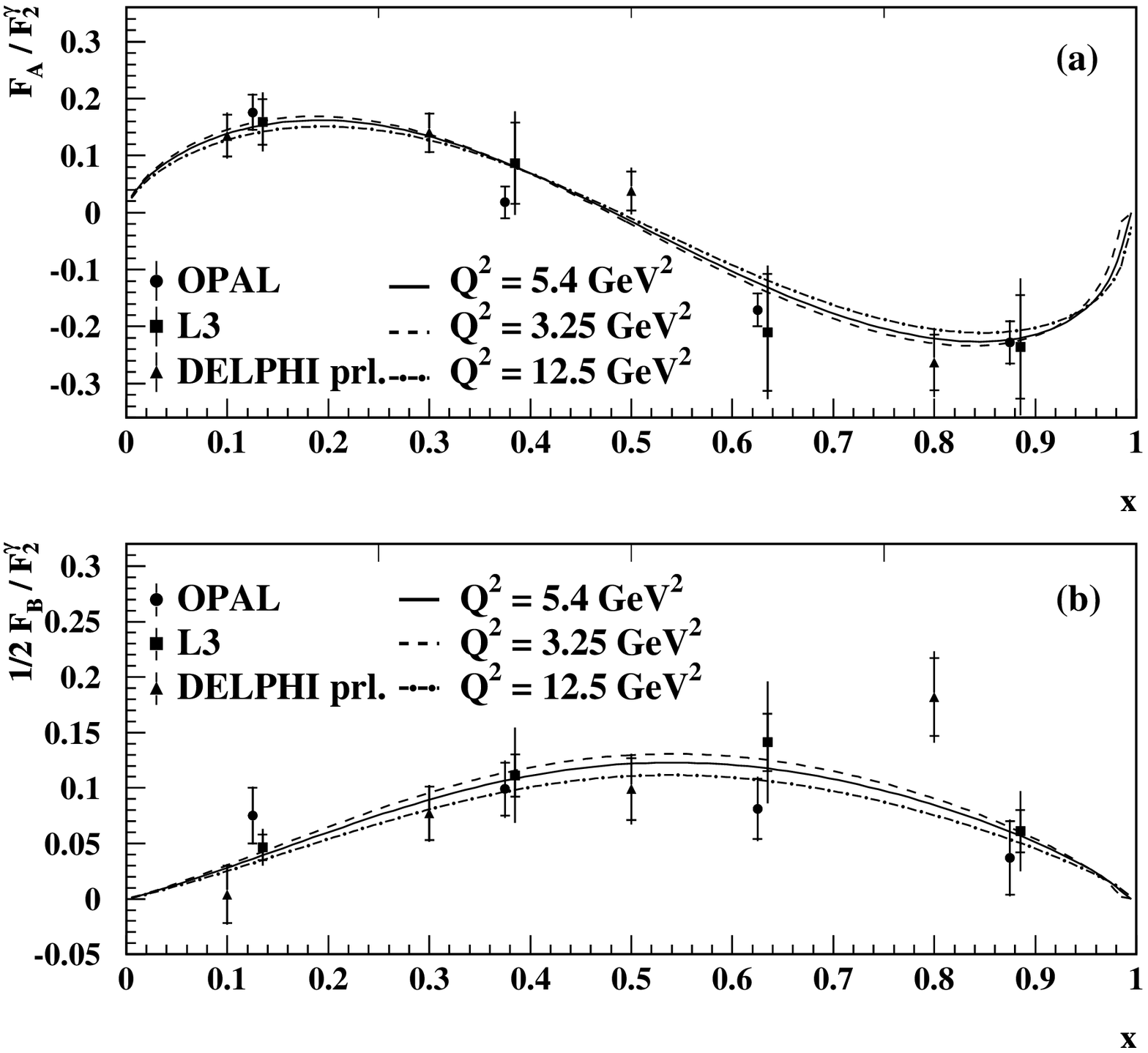}}
\mbox{}\vspace{-1.3cm}
\caption{
         The measurements of \faoft\ and 1/2\fboft.
        }\label{fig04}
\mbox{}\vspace{-1.3cm}
\end{center}
\end{figure}

 The structure functions \faqed\ and \fbqed\ are obtained
 from the measured \ftqed\ and a fit to the shape
 of the distribution of the azimuthal angle \az, which is the angle
 between the plane defined by the momentum vectors of the muons
 and the plane defined by the momentum vectors of the incoming and the
 deeply inelastically scattered electron.
 For small values of $y$, the \az\ distribution can be written as:
%
 \begin{equation}
 \frac{\der\,N}{\der\,\az} \sim 
 1 - A\cos\az + B\cos 2\az\,,
 \label{eqn:fit}
 \end{equation}
%
 with parameters $A=\faoft$ and $B=1/2\fboft$.
 The recent theoretical predictions~\cite{SEY-9801} which 
 take into account the important mass corrections up to \omfwq\
 are consistent with the results from L3~\cite{L3C-9801} and
 OPAL~\cite{OPALPR271} and with the preliminary results from 
 DELPHI~\cite{ZIN-9901}, as shown for the ratios in Figure~\ref{fig04}.
%
%
 Both \faqed\ and \fbqed\ are found to be significantly different 
 from zero.
%
%
\subsection{The cross-section for two exchanged virtual photons}
\label{sec:qedresvi}
 The cross-section for the exchange of two highly virtual photons 
 in the kinematical region under study can schematically
 be written as:
%
 \begin{eqnarray}
 \sigma &\sim&  \stt + \stl + \slt + \sll + \nonumber\\
        &    &  \frac{1}{2}\ttt\costph - 4\ttl\cosph\, .
\label{eqn:true}
\end{eqnarray}
%
 Here the total cross-sections \stt, \stl, \slt\ and \sll\ and the 
 interference terms \ttt\ and \ttl\ correspond to specific helicity states 
 of the photons (T=transverse and L=longitudinal), and \barph\ is the 
 angle between the electron scattering planes.
%
\begin{figure}[htb]\unitlength 1pt
\begin{center}
\mbox{}\vspace{-1.0cm}
{\includegraphics[width=1.0\linewidth]{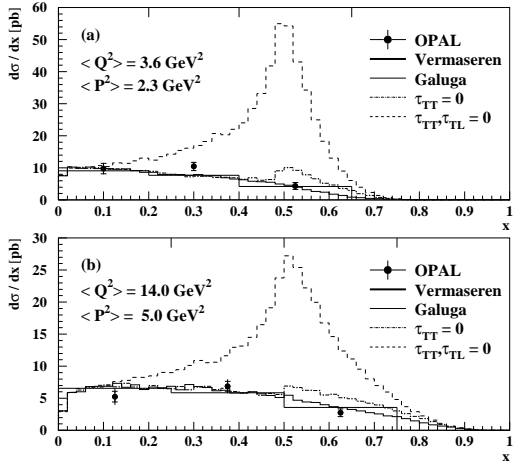}}
\mbox{}\vspace{-1.3cm}
\caption{
         The measurement of the differential QED cross-section \dsigdx\
         for highly virtual photons.
        }\label{fig05}
\mbox{}\vspace{-1.5cm}
\end{center}
\end{figure}
%
 There is good agreement between the measured \dsigdx\ from
 OPAL~\cite{OPALPR271} and the QED predictions using the Vermaseren 
 and the GALUGA Monte Carlo programs, provided all terms of the
 differential cross-section are taken into account.
 However, as apparent from Figure~\ref{fig05}, if either \ttt\ (dot-dash)
 or both \ttt\ and \ttl\ (dash) are neglected in the
 QED prediction as implemented in the GALUGA Monte Carlo, there is a
 clear disagreement between the data and the QED prediction.
 This measurement shows that both terms, \ttt\ and especially \ttl, 
 are present in the data in the kinematical region of the analysis,
 mainly at $x>0.1$, and that the corresponding contributions to the 
 cross-section are negative.
 \par
 As the kinematically accessible range in terms of \qsq\ and \psq\
 for the measurement of the QED and the QCD structure
 of the photon is the same, and given the size of the interference
 terms in the leptonic case, special care has to be taken when the
 measurements on the QCD structure are interpreted in terms
 of hadronic structure functions of virtual photons.
%
%
\section{The hadronic structure of the photon}
\label{sec:qcdres}
 The hadronic structure function \ft\ contains two contributions
 from the different appearances of the resolved photon, namely
 the point-like component stemming from the point-like coupling
 of the photon to quarks, and the hadron-like component originating 
 from the fluctuations of the photon into a hadronic state with the
 same quantum numbers as the photon. 
 The hadron-like part can be modelled by Vector Meson Dominance
 and obeys the same evolution equations as the structure function of
 an ordinary hadron like the proton.
 \par
%
\begin{figure}[htb]\unitlength 1pt
\begin{center}
\mbox{}\vspace{-1.0cm}
{\includegraphics[width=1.0\linewidth]{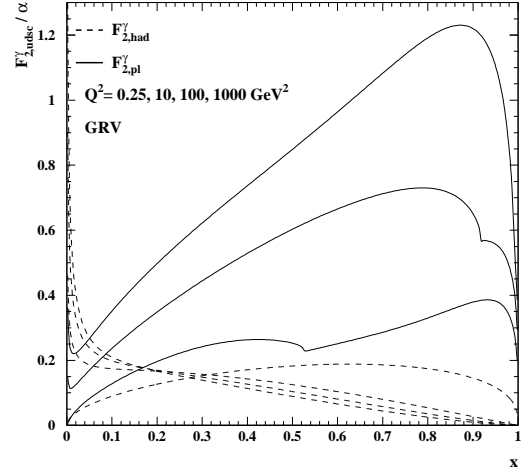}}
\mbox{}\vspace{-1.3cm}
\caption{
         The point-like (pl) and the hadron-like (had) contribution to
         \ft\ as predicted by the GRV parametrisation.
        }\label{fig06}
\mbox{}\vspace{-1.3cm}
\end{center}
\end{figure}
%
 The composition of these two components predicted by the GRV 
 parametrisation of \ft\ is shown in Figure~\ref{fig06}.
 With increasing \qsq\ the hadron-like contribution dies out at large
 $x$ and creates the steep rise at low $x$.
 In contrast, the point-like component increases with \qsq\ 
 at large $x$, the unique feature of \ft\ compared to \ftp.
 In the present investigations of \ft\
 both these features are investigated.
 The shape of \ft\ as a function of $x$ at fixed \qsq\ is studied with
 emphasis on the low-$x$ behaviour of \ft\ in comparison to \ftp, 
 and the evolution of \ft\ with \qsq\ is measured mainly at medium $x$.
%
%
\subsection{The description of the hadronic final state}
\label{sec:qcdreshad}
 The adequate description of the hadronic final state by the Monte Carlo
 models is very important for measurements of the photon structure.
 During the LEP2 workshop general purpose Monte Carlo programs became 
 available for deep inelastic electron-photon scattering.
 The first serious attempt to confront these models with the
 experimental data has been performed by OPAL~\cite{OPALPR185}.
 It was found, that none of the Monte Carlo programs available at that
 time was able to satisfactorily reproduce the data distributions.
 Therefore, the full spread of the predictions was included in the
 systematic error of the \ft\ measurement, which consequently suffered
 from large systematic errors.
 \par
 An important quantity is the flow of hadronic energy as 
 a function of the pseudorapidity, \flow.
 A significant fraction of the energy flow goes into the
 forward region of the detectors which are only equipped
 with electromagnetic calorimeters.
 Given this, the detectors are precise enough to disentangle various
 predictions in the central part of the detector, however they are
 not able to distinguish very well between models which produce
 different energy flow distributions in the forward region.
 \par
 The various Monte Carlo models produce significantly different hadronic
 energy flows, which leads to the fact that for a given value of $W$,
 the visible invariant mass \Wvis\ is rather different
 when using different Monte Carlo models, with the largest
 differences occuring at low values of $x$.
 \par
 In addition to the models discussed above the PHOJET and the
 TWOGAM Monte Carlo models were used in a structure function 
 analysis from L3~\cite{L3C-9803}, where the prediction of the hadronic
 energy flow for these two models was compared to the data for the \qsq\
 range $1.2-9$~\gevsq.
 Again, these two models, although closer to the data than the
 HERWIG5.8d
 and PYTHIA predictions in the case of the OPAL analysis, do not
 accurately account for the features observed in the data distributions.
 \par
 Several methods were investigated to reduce the dependence on the
 Monte Carlo models:\newline
 $\bullet$
 Motivated by the observation made in photoproduction studies at 
 HERA~\cite{ZEU-9501}, the HERWIG5.9 model was 
 tuned~\cite{LAU-9701CAR-9701} by changing 
 the distribution of \kt, the intrinsic transverse momentum of the
 quarks inside the photon from a Gaussian to a power-law 
 behaviour of the form $\der\ktsq/(\ktsq+\kn^2)$.
 In the latest version the upper limit of \kt\ is dynamically adjusted on an
 event by event basis.
 \par
 $\bullet$
 The longitudinal momentum of the photon-photon system is unknown,
 but the transverse momentum is well constrained  by measuring
 the transverse momentum of the scattered electron.
 This fact is used to replace a part of the measurement of the
 hadronic system by quantities obtained from the scattered electron,
 and thereby a part of the uncertainty of the measurement
 of the hadronic final state can be eliminated.
 The distribution of the invariant mass reconstructed in this scheme
 by L3~\cite{L3C-9803} is closer to the $W$ distribution than the
 \Wvis\ distribution, but still the agreement with $W$ is not very good.
 \par
 $\bullet$
 Another way of reducing the model dependence is to perform
 the unfolding in two dimensions.
 Preliminary results from the ALEPH~\cite{BOE-9901} and 
 OPAL~\cite{CLA-9901} experiments show that this indeed reduces the
 systematic uncertainty on the \ft\ measurement.
 \par
 $\bullet$
 In order to establish a consistent picture a combined effort by the
 ALEPH, L3 and OPAL collaborations and the LEP  Two-Photon Working
 Group has been undertaken~\cite{FIN-9901}.
 The data of the experiments have been analysed in two regions of
 \qsq, $1.2-6.3$~\gevsq\ and $6-30$~\gevsq, using identical cuts
 and also identical Monte Carlo events passed through the respective
 programs of the individual experiments to simulate the detector response.
 The data distributions are corrected to the hadron level, allowing
 for a direct comparison to the predictions of the Monte Carlo models.
 It was found, that for large regions in most of the distributions studied,
 the results of the different experiments are closer to each other
 than the sizable differences which are found between the data and the
 models.
 \par
 From the discussion above it is clear, that the error on the
 measurement of \ft\ will vary strongly with the Monte Carlo
 models chosen to obtain the systematic uncertainty.
 However, given the improved understanding of the shortcomings
 and the combined effort on improving on the Monte Carlo
 description of the data, it is likely that the error on \ft\
 will shrink considerably in future measurements.
%
%
\subsection{The structure of quasi-real photons}
\label{sec:qcdresf2}
 Many measurements of the hadronic structure function \ft\ have been
 performed at several electron-positron colliders.
 In this review
 the interpretation of the data will be based on the published results,
 and on those preliminary results from the LEP experiments,
 which are based on data which have not yet been published.
 For the preliminary results~\cite{CLA-9901} which are meant to replace
 a published measurement in the near future the published result will
 be shown until the new result is finalised.
 \par
\begin{figure}[htb]\unitlength 1pt
\begin{center}
\mbox{}\vspace{-1.5cm}
{\includegraphics[width=1.0\linewidth]{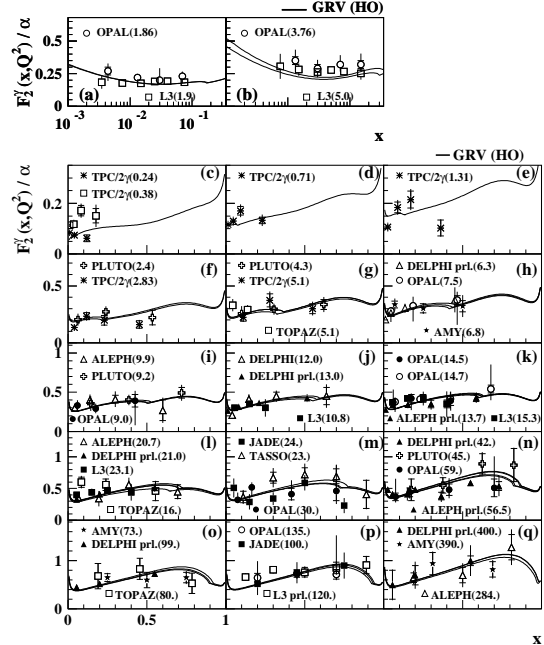}}
\mbox{}\vspace{-1.3cm}
\caption{
         A summary of measurements of \ft.
        }\label{fig07}
\mbox{}\vspace{-1.3cm}
\end{center}
\end{figure}

 The range in \qzm\ covered by the various experiments is
 $0.24-400$~\gevsq, which is impressive given the small cross-section
 of the process.
 In Figure~\ref{fig07} the published results from
 ALEPH~\cite{ALE-9901}
 AMY~\cite{AMY-9501AMY-9701},
 DELPHI~\cite{DEL-9601},
 JADE~\cite{JAD-8401},
 L3~\cite{L3C-9803,L3C-9804},
 OPAL~\cite{OPALPR185,OPALPR207,OPALPR213},
 PLUTO~\cite{PLU-8401PLU-8701},
 TASSO~\cite{TAS-8601},
 TPC/2$\gamma$~\cite{TPC-8701}
 and
 TOPAZ~\cite{TOP-9402}
 are shown together with additional preliminary results from
 ALEPH~\cite{BOE-9901}, DELPHI~\cite{TIA-9701} and L3~\cite{ERN-9901}.
 \par
 The precision of the measurements of \ft\ which have been performed
 at \epem\ centre-of-mass energies below the mass of the $Z$ boson
 is mainly limited by the statistical error and, due to the simple 
 assumptions made on the hadronic final state, the systematic errors
 are small, but in light of the discussion above, they maybe
 underestimated.
 Some of the data show quite unexpected features. For example, the \ft\
 as obtained from TPC/2$\gamma$ shows an unexpected shape at low values
 of $x$, and also the results from TOPAZ rise very fast towards low
 values of $x$. In addition there is a clear disagreement between
 the TASSO and JADE data at $\qzm= 23-24$~\gevsq.
 \par
%
%
 The measurement of \ft\ has attracted a lot of interest at LEP over
 the last years. The LEP Collaborations have measured \ft\ in the 
 range $0.002<x$ \lsim\ 1 and $1.86<\qzm<400$ \gevsq.
 As seen from Figure~\ref{fig07} the measurements obtained at LEP1 
 energies (open symbols) are consistent with those obtained at LEP2 
 energies (closed symbols).
 \par
 The comparison to the GRV higher order prediction of \ft\ shows an
 overall agreement, but also some regions where the prediction does not
 so well coincide with the data.
 This large amount of data, which partly is rather precise, gives the
 possibility to study the consistency of the predictions with the
 data.
 The quality of the agreement is evaluated by a simple \chiq\ method
 based on:
%
 \begin{equation}
  \chiq = \sum_i \left(
                \frac{\fti-\langle\ft(x,\qzm,0)\rangle}{\sigma_i}
                \right)^2\, ,
 \label{eqn:chiq}
 \end{equation}
%
 where the sum runs over all measurements in Figure~\ref{fig07} with
 $\qzm>\qnsq$, where \qnsq\ is the starting scale of the evolution for
 the respective parametrisation of \ft.
 The term \fti\ denotes the measured value of \ft\ in the $i^{th}$ bin and
 $\sigma_i$ is its total error. The theoretical expectation is
 approximated by the average \ft\ in that bin in $x$ for $\qsq=\qzm$ and
 for $\psq=0$, abbreviated by $\langle\ft(x,\qzm,0)\rangle$.
 The procedure does not take into account the 
 correlation of errors between the data points.
 A more accurate analysis would require to study in detail the
 correlation between the results within one experiment,
 but even more difficult, the correlation between
 the results from different experiments, a major task which is beyond
 the scope of the comparison presented here.
 The predictions used in the comparison are the WHIT
 parametrisations, which are the most recent parametrisations based
 on purely phenomenological fits to the data, and the GRV and SaS1D
 predictions which use some theoretical prejudice to construct
 the hadron-like part of \ft\ at \qnsq.
 The values of \chidof\ found are: GRV LO (1.55),
 GRV HO (1.64), GRSchienbein LO (1.58), SaS1D (1.81),
 SaS1M (1.50), for $165/161$ data points with $\qzm >0.38/0.71$~\gevsq,
 and SaS2D (0.97), SaS2M (1.01),
 WHIT1 (1.10), WHIT2 (3.27), WHIT3 (5.37), WHIT4 (5.78), WHIT5 (18.66),
 WHIT6 (28.29), for 132 data points with $\qzm >4$~\gevsq.
 The parametrisations describe the AMY, JADE, PLUTO and TASSO data, and 
 they all disfavour the TPC/2$\gamma$ results.
 The WHIT parametrisations predict a faster rise at low-$x$
 than the GRV, GRSc and the SaS parametrisations.
 Therefore, the agreement with the TOPAZ data is satisfactory for
 the WHIT parametrisations, whereas the GRV, GRSc and the SaS1
 parametrisations yield values of \chidof\ of around 2,
 and the SaS2 parametrisations lie somewhere between these extremes.
 For the same reason the WHIT parametrisations fail to describe the
 ALEPH and DELPHI data which tend to be low at low values of $x$,
 thereby leading to large \chidof\ for the WHIT parametrisations,
 especially for the sets WHIT4-6 which use a larger gluon distribution
 functions. The only acceptable agreement is achieved by using the
 set WHIT1.
 The OPAL results tend to be high at low values of $x$ and also
 they have larger errors, therefore only the extreme cases
 WHIT5-6 lead to unacceptable values of \chidof.
 The quoted uncertainties by L3 are very small and the
 results tend to be high for low values of \qsq.
 Consequently, none of the parametrisations which are valid
 below $\qsq=4$~\gevsq\ is able to describe the L3 data and all lead
 to large values of \chidof.
 For $\qsq>4$~\gevsq\ the agreement improves but the values of \chidof\
 are still too large, besides for GRSc.
 For the parametrisations valid for $\qsq>4$~\gevsq\ the best agreement
 with the L3 data is obtained for WHIT1.
 This comparison shows that already at the present level of
 accuracy the measurements of \ft\ are precise enough to constrain
 the parametrisations and to discard those which predict
 a fast rise at low $x$ driven by large gluon distribution functions.
 \par
\begin{figure}[htb]\unitlength 1pt
\begin{center}
\mbox{}\vspace{-1.7cm}
{\includegraphics[width=1.0\linewidth]{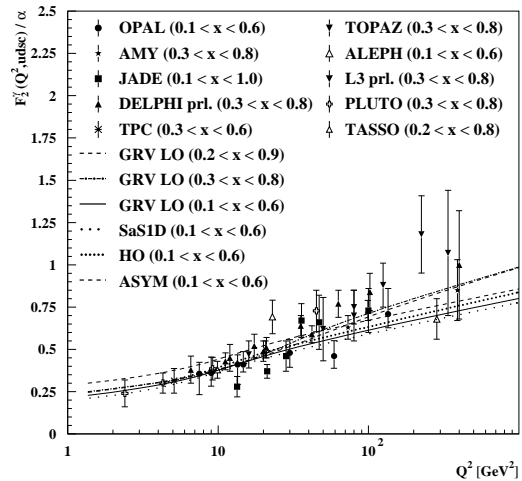}}
\mbox{}\vspace{-1.3cm}
\caption{
         The \qsq\ evolution of \ft\ at medium $x$.
        }\label{fig08}
\mbox{}\vspace{-1.3cm}
\end{center}
\end{figure}

 The second topic which is extensively studied using the large lever
 arm in \qsq, is the evolution of \ft\ with \qsq.
 A collection of all available measurements of the evolution of
 \ft\ at medium $x$  for four active flavours is shown 
 in Figure~\ref{fig08}.
 For PLUTO the average \ftc\ in the range $0.2<x<0.8$ for the \qzm\ values
 of the analyses has been added to the published three flavour result.
 \par
 Unfortunately the different experiments quote their results for different
 ranges in $x$ which makes the comparison more difficult because
 the predictions for the various ranges in $x$ start to be significantly
 different for $\qsq>100$~\gevsq, as can be seen in
 Figure~\ref{fig08}.
 The measurements are consistent with each other and a clear rise of
 \ft\ with \qsq\ is observed.
 It is an interesting fact that this rise can be described
 reasonably well (${\mathcal O}(15\%)$ accuracy) by the 
 leading order augmented asymptotic prediction detailed in~\cite{OPALPR207},
 which uses the asymptotic solution~\cite{WIT-7701}
 for \ft\ for the light flavour contribution as predicted by
 perturbative QCD for $\almz=0.128$.
 \par
%
\begin{figure}[htb]\unitlength 1pt
\begin{center}
\mbox{}\vspace{-1.3cm}
{\includegraphics[width=1.0\linewidth]{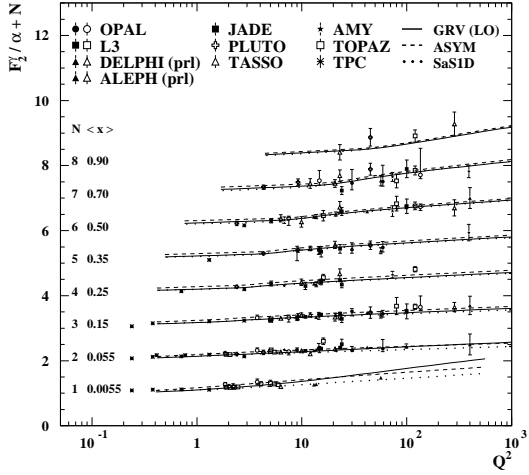}}
\mbox{}\vspace{-1.3cm}
\caption{
         Summary of the measurements of the \qsq\ evolution of \ft.
        }\label{fig09}
\mbox{}\vspace{-1.3cm}
\end{center}
\end{figure}

%
 In order to look at the variation of the scaling violation as a function
 of $x$ the data from Figure~\ref{fig07} are displayed differently
 in Figure~\ref{fig09}.
 The data are shown as a function of \qsq, divided in bins of $x$,
 with average values as shown in the figure.
 Each individual measurement is attributed to the bin with the closest
 average value in $x$ used.
 To separate the measurements from each other an integer value, N,
 counting the bin number is added to the measured \ft.
 The theoretical prediction is the average \ft\ in each bin.
 The \qsq\ ranges used for the predictions are the maximum ranges
 possible for $1<W<250$~\gev\ and $\qnsq<\qsq<1000$~\gevsq.
 The general trend of the data is followed by the predictions
 of the augmented asymptotic solution, and
 the GRV and SaS1D leading order parametrisations of \ft,
 however, differences are seen which were discussed above.
 \par
%
%
%
%
\subsection{The hadronic structure of virtual photons}
\label{sec:qcdresvirt}
 The structure functions of virtual photons can be determined
 in the region $\qsq\gg\psq$ by measuring the cross-sections for events
 where both electrons are observed.
 An effective structure function
 $\feff\propto \stt + \stl + \slt + \sll + 1/2\ttt\costph -  4\ttl\cosph$
 can be measured by experiments, however, to relate \feff\ to the structure
 functions \ft\ and \fl\ further assumptions are needed.
 By assuming that the interference terms do not
 contribute, that \sll\ is negligible and also using $\stl=\slt$,
 the relation $\feff = \ft + 3/2 \fl$ is derived.
 Due to the \psq\ suppression of the cross-section these measurements
 suffer from low statistics.
 \par
 The first measurement of this type performed by PLUTO~\cite{PLU-8405},
 for $\pzm=0.35$~\gevsq\ and $\qzm=5$~\gevsq, has been compared 
 in~\cite{GLU-9501} to the
 theoretical predictions from the GRS parametrisations.
 The best description of the data is obtained using the next-to-leading
 order result including a non-perturbative input at the starting scale
 of the evolution.
 If the hadron-like input is neglected, the prediction is consistently
 lower than the data, but still consistent with it, within the experimental
 errors.
 Also the prediction obtained by calculating \feff\ solely from the
 box diagram is still consistent with the data although it is the lowest
 at high values of $x$.
 The evolution of \feff\ with \psq\ has been studied as well,
 including the result of \ft\ for the quasi-real photon.
 The data suggest a slow decrease with increasing \psq, but
 they are also consistent with a constant behaviour.
 Also for the \psq\ evolution of \feff\ the full next-to-leading order 
 prediction gives the best description of the data and the purely
 perturbative prediction is at the low end.
 \par
 A similar measurement has been presented by L3~\cite{ERN-9901}.
 The average virtualities for the L3 result are $\qzm = 120$~\gevsq\
 and $\pzm = 3.7$~\gevsq, thereby ensuring $\qsq\gg\psq\gg\Lambda^2$.
 As in the case of PLUTO the QPM result is too low compared to the
 data. Taking only \ft\ as calculated from the GRS parametrisation of the
 parton distribution functions of the photon
 gets closer to the data, and the best description is found
 if the contribution of \fl\ is added to this, based on the prediction
 of the QPM.
 The data show a faster rise with $x$ than any of the predictions,
 however with large errors for increasing $x$, which are mainly due
 to the low statistics available.
 \par
 The QPM prediction of the \psq\ evolution of \feff\ is consistent in
 shape with the data, but the predicted \feff\ is too low.
 However, the main difference comes from \ft\ at $\psq=0$, which
 is not described by the quark parton model for $x<0.4$,
 where the hadron-like component is expected to be largest.
 But in this region the data are even higher than the predictions
 of all parametrisations of \ft\ which contain a hadron-like
 contribution.
 The measurement at $\psq>0$ cannot rule out the quark parton model
 prediction, although the data are consistently higher.
 The ratio of \qzm/\pzm\ is similar for the PLUTO and L3 measurements,
 leading to values for $\ln(\qzm/\pzm)$ of 2.6 and 3.5 respectively.
 This enables to compare the \qsq\ evolution of the two measurements.
 The evolution is consistent with the expectation of the quark parton
 model for $\ln(\qzm/\pzm)=3$, and using the range $0.05<x<0.98$.
 \par
 In summary a consistent picture is found for the effective structure
 function of the virtual photon between the PLUTO and L3 data and 
 the general features of both measurements
 are described by the next-to-leading order predictions.
 However, the data do not constrain the predictions very strongly and
 for more detailed comparisons to be made the full statistics of the LEP2
 programme has to be explored.
 \par
 If both photons have similar virtualities the application of the photon
 structure function picture is no longer applicable and the data are
 interpreted in terms of the differential cross-section.
 Due to the large virtualities the cross-section is small and
 large integrated luminosities are needed to precisely measure it.
 The main interest is the investigation of the hadronic structure
 of the interaction of two virtual photons.
 However,
 the interest in performing these measurements increased considerably
 in the last years, because calculations in the framework of the
 leading order BFKL evolution equation, which sums $\ln(1/x)$
 contributions, predicted a large 
 cross-section~\cite{BAR-9601BAR-9701BRO-9701BRO-9702}.
 The predicted cross-section is so large that already measurements
 with low statistics are able to decide whether the BFKL picture is
 in agreement with the measurements.
 Recently theoretical progress has been made~\cite{FAD-9801CAM-9801} to
 also include next-to-leading order pieces in the 
 BFKL calculations~\cite{BRO-9901}.
 Large negative corrections to the leading order results were 
 found, e.g.~\cite{BON-9801}, consequently, there is
 some doubt about the perturbative stability of the calculation.
 The theoretical development is underway and this should be kept 
 in mind in comparisons to the BFKL predictions.
 The most suitable region for the comparison is
 $\wsq\gg\qsq\approx\psq\gg\Lambda^2$, which ensures
 similar photon virtualities and large values of 1/$x$.
 These requirements strongly reduce the available statistics,
 therefore compromises have to be made in this comparisons.
 \par
 The first measurement of this type was performed by
 L3~\cite{L3C-9903} using data at $\ssee=91$ and $183$~\gev, and 
 additional preliminary results were 
 reported~\cite{ACH-9901} for data taken at $\ssee=189$~\gev.
 The average photon virtualities \qzm,\pzm\ are 3.5, 14 and
 14.5~\gevsq\ respectively. The $W$ ranges used are
 $2-30/5-70/5-75$~\gev\ for the three centre-of-mass energies, which means the
 lowest values of $x$ probed are about $2-3\cdot 10^{-3}$.
 The differential cross-section as a function of
 $Y=\ln(\wsq/\sqrt{\qsq\psq})$ is described by the TWOGAM Monte Carlo
 at $\ssee=91$ and $183$~\gev.
 The PHOJET model gives an adequate description at $\ssee=183$ and
 $189$~\gev, whereas it fails to describe the data at
 $\ssee=91$~\gev, probably due to the low cut in $W$ applied.
 The prediction of the QPM is found to be too low at all energies.
 The cross-sections predicted by the BFKL calculation are much
 higher than what is observed and are strongly disfavoured by the data.
 \par
 A similar analysis was presented by OPAL~\cite{PRY-9901},
 based on data at $\ssee=189$~\gev, with average photon
 virtualities of about $10$~\gevsq, and for $W>5$~\gev.
 The differential cross-section is obtained as functions of $W$, $x$ 
 and \qsq, for $W>5$~\gev\ and for energies of the scattered electrons
 larger than 65~\gev, and polar angles in the range $34-55$~mrad.
 The measured cross-section in this region is
 $0.32 \pm 0.05 (stat)\,^{+0.04}_{-0.05} (sys)$~pb,
 compared to the predicted cross-sections of $0.17$~pb for PHOJET and
 $2.2 / 0.26$~pb based on the BFKL calculation in leading/'higher order'.
 Also for OPAL the data are described by the PHOJET model and there
 is no room for large additional contributions.
 The precision of the results is limited by statistics and they can
 be improved by using the full statistics of the LEP2 programme.
%
%
\section{Conclusions}
\label{sec:concl}
 Many new results on the QED and QCD structure of the photon have been 
 obtained in the last years.
 \par
 The QED predictions of the structure of the photon are found to be in
 good agreement with all experimental results.
 \par
 For the measurement of the hadronic structure function \ft\
 considerable progress has been made concerning the problem of the
 inaccurate modelling of the hadronic final state by the Monte Carlo models.
 New results on the low-$x$ behaviour of \ft\ have been presented
 and by now the evolution of \ft\ with \qsq\ is probed up to 
 $\qsq=400$~\gevsq.
 \par
 Interesting studies on the hadronic structure of the exchange of two 
 virtual photons have been performed. The effective structure function 
 has been measured at LEP and its behaviour can be described by 
 the next-to-leading order prediction.
 The cross-section for the exchange of two highly virtual photons can
 be described with conventional models, whereas
 the present BFKL predictions are strongly disfavoured by the data.
 \newline
%
%
 {\bf Acknowledgement:}\\
 I wish to thank the organisers of this interesting conference
 for the fruitful atmosphere they created throughout the meeting. 
%
%

%

\begin{thebibliography}{10}
\bibitem{NISIUS}
 R.~Nisius,
 in {\em DIS99 Conf., Zeuthen}, World Scientific, 1999, hep-ex/9905059;
 {\em DIS98 Conf., Brussels}, eds.
 G.~Coremans and R.~Roosen, pages 194--198, World Scientific, 1998;
 and {\em ICHEP97 Conf., Jerusalem}, 1997, hep-ex/9712012.
\bibitem{PLU-8102}
 PLUTO Collab., C.~Berger et~al., Phys. Lett. {\bf 107B}, 168--172 (1981).
\bibitem{CEL-8301}
 CELLO Collab., {H.J. Behrend} et~al., Phys. Lett. {\bf 126B}, 384--390 (1983).
\bibitem{DEL-9601}
 DELPHI Collab., P.~Abreu et~al., Z. Phys. {\bf C69}, 223--234 (1996).
\bibitem{L3C-9801}
 L3 Collab., M.~Acciarri et~al., Phys. Lett. {\bf B438}, 363--378 (1998).
\bibitem{OPALPR271}
 OPAL Collab., G.~Abbiendi et~al., CERN-EP/99-010.
\bibitem{PLU-8501}
 PLUTO Collab., C.~Berger et~al., Z. Phys. {\bf C27}, 249--256 (1985).
\bibitem{TPC-8401}
 TPC/2$\gamma$ Collab., {M.P. Cain} et~al., Phys. Lett. {\bf 147B}, 
 232--236 (1984).
\bibitem{BRE-9701}
 ALEPH Collab., C.~Brew,
 in {\em Photon '97, Egmond aan Zee}, eds.
 A.~Buijs and {F.C. Ern{\'e}}, pages 21--26, World Scientific, 1998.
\bibitem{ZIN-9901}
 DELPHI Collab., A.~Zintchenko, these proceedings.
\bibitem{SEY-9801}
 R.~Nisius and {M.H. Seymour}, Phys. Lett. {\bf B452}, 409--413 (1999).
\bibitem{OPALPR185}
 OPAL Collab., K.~Ackerstaff et~al., Z. Phys. {\bf C74}, 33--48 (1997).
\bibitem{L3C-9803}
 L3 Collab., M.~Acciarri et~al., Phys. Lett. {\bf B436}, 403--416 (1998).
\bibitem{ZEU-9501}
 ZEUS Collab., M.~Derrick et~al., Phys. Lett. {\bf B354}, 163--177 (1995).
\bibitem{LAU-9701CAR-9701}
 {J.A. Lauber}, L.~L{\"o}nnblad, and {M.H. Seymour},
 in {\em Photon '97, Egmond aan Zee}, eds.
 A.~Buijs and {F.C. Ern{\'e}}, pages 52--56, World Scientific, 1998;
 S.~Cartwright, {M.H. Seymour}, et~al., J. Phys. G {\bf 24}, 457--481 (1998).
\bibitem{BOE-9901}
 ALEPH Collab., {A.~B\"ohrer}, these proceedings.
\bibitem{CLA-9901}
 OPAL Collab., E.~Clay, these proceedings.
\bibitem{FIN-9901}
 ALEPH, L3 and OPAL Collabs., A.~Finch, these proceedings.
\bibitem{ALE-9901}
 ALEPH Collab., D.~Buskulic et~al., CERN-EP/99-063.
\bibitem{AMY-9501AMY-9701}
 AMY Collab., {S.K. Sahu} et~al., Phys. Lett. {\bf B346}, 208--216 (1995);
 T.~Kojima et~al., Phys. Lett. {\bf B400}, 395--400 (1997).
\bibitem{JAD-8401}
 JADE Collab., W.~Bartel et~al., Z. Phys. {\bf C24}, 231--245 (1984).
\bibitem{L3C-9804}
 L3 Collab., M.~Acciarri et~al., Phys. Lett. {\bf B447}, 147--156 (1999).
\bibitem{OPALPR207}
 OPAL Collab., K.~Ackerstaff et~al., Phys. Lett. {\bf B411}, 387--401 (1997).
\bibitem{OPALPR213}
 OPAL Collab., K.~Ackerstaff et~al., Phys. Lett. {\bf B412}, 225--234 (1997).
\bibitem{PLU-8401PLU-8701}
 PLUTO Collab., C.~Berger et~al., Phys. Lett. {\bf 142B}, 111--118 (1984);
 Nucl. Phys. {\bf B281}, 365--380 (1987).
\bibitem{TAS-8601}
 TASSO Collab., M.~Althoff et~al., Z. Phys. {\bf C31}, 527--535 (1986).
\bibitem{TPC-8701}
 TPC/2$\gamma$ Collab., H.~Aihara et~al., Z. Phys. {\bf C34}, 1--13 (1987).
\bibitem{TOP-9402}
 TOPAZ Collab., K.~Muramatsu et~al., Phys. Lett. {\bf B332}, 477--487 (1994).
\bibitem{TIA-9701}
 DELPHI Collab., I.~Tyapkin, in {\em Photon '97, Egmond aan Zee}, eds.
 A.~Buijs and {F.C. Ern{\'e}}, pages 26--30, World Scientific, 1998.
\bibitem{ERN-9901}
 L3 Collab., {F.C. Ern{\'e}}, these proceedings.
\bibitem{WIT-7701}
 E.~Witten, Nucl. Phys. {\bf B120}, 189--202 (1977).
\bibitem{PLU-8405}
 PLUTO Collab., C.~Berger et~al., Phys. Lett. {\bf 142B}, 119--124 (1984).
\bibitem{GLU-9501}
 M.~Gl{\"u}ck, E.~Reya, and M.~Stratmann, 
 Phys. Rev. {\bf D51}, 3220--3229 (1995).
\bibitem{BAR-9601BAR-9701BRO-9701BRO-9702}
 J.~Bartels, A.~{De Roeck}, and H.~Lotter,
 Phys. Lett. {\bf B389}, 742--748 (1996);
 J.~Bartels, A.~{De Roeck}, C.~Ewerz, and H.~Lotter, hep-ph/9710500;
 {S.J. Brodsky}, F.~Hautmann, and {D.E. Soper},
 Phys. Rev. {\bf D56}, 6957--6979 (1997);
 Phys. Rev. Lett. {\bf 78}, 803--806, Erratum--ibid {\bf 79} 3544 (1997).
\bibitem{FAD-9801CAM-9801}
 {V.S. Fadin} and {L.N. Lipatov}, Phys. Lett. {\bf B429}, 127--134 (1998);
 G.~Camici and M.~Ciafaloni, Phys. Lett. {\bf B430}, 349--354 (1998).
\bibitem{BRO-9901}
 {S.J. Brodsky} et~al., hep-ph/9901229.
\bibitem{BON-9801}
 M.~Boonekamp, A.~{De Roeck}, C.~Royon, and S.~Wallon, hep-ph/9812523.
\bibitem{L3C-9903}
 L3 Collab,  M.~Acciarri et~al., Phys. Lett. {\bf B453}, 333--342 (1999).
\bibitem{ACH-9901}
 L3 Collab., P.~Achard, these proceedings.
\bibitem{PRY-9901}
 OPAL Collab., {M. Przybycie\'n}, these proceedings.
\end{thebibliography}
\end{document}